\documentclass[aps,prb,twocolumn,showpacs,groupedaddress]{revtex4}
\usepackage{graphicx}
\usepackage{amssymb}
\usepackage{amsmath}
\usepackage{xcolor}

\usepackage{epsf}
\begin{document}
\preprint{\today}
\title{Delay-Time and Thermopower Distributions \\ 
at the Spectrum Edges of a Chaotic Scatterer}
\author{Adel Abbout}
\altaffiliation[Present Address: ]{Laboratoire CRISMAT, CNRS UMR 6508, 6 boulevard Mar\'echal Juin, F-14050, Caen, France}
\affiliation{Service de Physique de l'\'Etat Condens\'e (CNRS URA 2464), 
IRAMIS/SPEC, CEA Saclay, 91191 Gif-sur-Yvette, France}

\author{Genevi\`eve Fleury}
\affiliation{Service de Physique de l'\'Etat Condens\'e (CNRS URA 2464), 
IRAMIS/SPEC, CEA Saclay, 91191 Gif-sur-Yvette, France}

\author{Khandker Muttalib}
\affiliation{Department of Physics, University of Florida, Gainesville, FL 32611-8440, USA}

\author{Jean-Louis Pichard}
\affiliation{Service de Physique de l'\'Etat Condens\'e (CNRS URA 2464), 
IRAMIS/SPEC, CEA Saclay, 91191 Gif-sur-Yvette, France}

\begin{abstract}
We study chaotic scattering outside the wide band limit, as the Fermi energy $E_F$ approaches the 
band edges $E_B$ of a one-dimensional lattice embedding a scattering region of $M$ 
sites. We show that the delay-time and thermopower distributions differ near the edges 
from the universal expressions valid in the bulk. To obtain the asymptotic universal forms of these 
edge distributions, one must keep constant the energy distance $E_F-E_B$ measured in units of the same 
energy scale proportional to $\propto M^{-1/3}$ which is used for rescaling the energy level spacings at the spectrum 
edges of large Gaussian matrices. In particular the delay-time and the thermopower have the same universal 
edge distributions for arbitrary $M$ as those for an $M=2$ scatterer, which we obtain analytically.
\end{abstract}
\maketitle

\section{Introduction}
Since nano-engineering makes it possible to fabricate devices which can be used for harvesting energy (Seebeck effect) 
or for cooling (Peltier effect) at the nanoscales, it becomes necessary to study thermoelectric conversion in the quantum 
limit, where mesoscopic fluctuations become important. To illustrate that point, let us consider a chaotic quantum dot 
created by local depletion of a two-dimensional electron gas and connected to two leads by two quantum point contacts having 
a few opened channels. The thermopower ${\mathcal S}_k$ of the dot is defined as the ratio 
$-\Delta V / \Delta {\mathcal T}$ of a (small) voltage and temperature difference applied over the dot at zero electrical current. 
It has been measured~\cite{molenkamp} in chaotic dots of submicrometer size with two opened channels per contact, the electrons 
on one side of the dot being heated at $ \approx 50\, \mathrm{mK}$ above a sub-Kelvin lattice temperature. Mesoscopic fluctuations $\approx 20 
\,\mu \mathrm{V/K}$ around a zero average value have been observed either when the shape of the dot is changed (deformation of the order of the 
Fermi wavelength) or when an applied magnetic flux is varied (by about half a flux quantum). This example shows us that the thermoelectric 
conversion is associated with a large mesoscopic fluctuation and illustrates the necessity to determine the full distribution, and not 
only the average value, when one studies the thermopower of phase coherent nanostructures. \\
\indent Scattering theory, combined with random matrix theory (RMT) allows one to obtain~\cite{vlsb} such a distribution for a chaotic dot as long as 
it behaves as an elastic scatterer. Using the Landauer-Buttiker formalism, it is straightforward to rederive the Cutler-Mott 
formula~\cite{sivan-imry,van-houten,lunde}: 
\begin{equation}
{\mathcal S}_k=\frac{1}{e{\mathcal T}}\frac{\int dE (E-E_F) T(E) df/dE}{\int dE\, T(E) df/dE}
\label{cutler-mott}
\end{equation}
which gives the thermopower ${\mathcal S}_k$ (at temperature ${\mathcal T}$ and Fermi energy $E_F$) in terms of the function 
$T(E)$ giving the scatterer transmission as a function of the electron energy $E$. Here, $f$ is the Fermi-Dirac function and $e$ the electron 
charge. If the temperature ${\mathcal T}$ is smaller than the energy scale over which $T(E)$ fluctuates, the Wiedemann-Franz 
law is satisfied~\cite{stone-vavilov,bosisio}, Sommerfeld expansions can be made~\cite{lunde} and the thermopower ${\mathcal S}_k$ 
(in units of $\pi^2k_B^2{\mathcal T}/3e$, $k_B$ being the Boltzmann constant) reads 
\begin{equation}
S_k=\frac{1}{T}\frac{dT}{dE}
\label{cutler-mott-sommerfeld}
\end{equation}
where $T$ and $dT/dE$ are to be evaluated at $E=E_F$ only. In this low ${\mathcal T}$-limit, to obtain the distribution of $S_k$ requires to 
know the distribution of not only the scattering matrix $S$ but also of its energy derivative at $E_F$, which can 
be obtained from the Wigner-Smith time-delay matrix \cite{smith,bfb1,bfb2} $Q \equiv -i \hbar S^{-1}\partial S/ \partial E$. The 
distributions of $Q$, and hence of $S_k$, have been previously obtained~\cite{vlsb} from a RMT description valid for chaotic cavities, 
but only in the wide band limit (WBL), restricted to the \textit{bulk} of a wide conduction band. In that case, it is the bulk of 
the cavity spectrum which is probed at $E_F$. In the microscopic model which we study exactly, we consider not only the bulk 
of the conduction band, but also its \textit{edges}. In this case, it is not the bulk of the spectrum, but its edges which are probed at 
$E_F$ if one wants to keep the scattering chaotic. This is particularly interesting, since the universality near the edges can be different 
from that in the bulk, as is well-known from the Tracy-Widom vs Wigner level distributions~\cite{tracy-widom1,tracy-widom2,forrester}. We 
show in this paper that the distributions of $Q$ and $S_k$ become different near the edges, and give rise to a new asymptotic universality 
for the time-delay matrix $Q$ and for the thermopower $S_k$ when the Tracy-Widom scaling is adopted. 

While the observation of the distribution of $S_k$ requires very low temperatures~\cite{molenkamp}, the observation of the edge distribution of 
$Q$ can be done at room temperature if one uses chaotic cavities~\cite{doron} and waves~\cite{tiggelen,schanze} (electromagnetic, acoustic, $\ldots$) 
other than electrons. $Q$ yields information \cite{smith} on the time that a particle is delayed in a scattering region and gives also the ac-response of 
mesoscopic scatterers~\cite{gmb,bb,rie}. Its distribution has been studied in particular for wave reflection from a long random potential in the 
one-dimensional (1D) localized limit : Remarkably, a departure from the universal distribution valid in the bulk was found in the low energy 
limit~\cite{texier-comtet} of a continuum model and near the band edges~\cite{kumar} of a 1D disordered tight-binding model. In this paper, we study 
the departure from the bulk distributions of $S_k$ and $Q$, which occurs at the band edges of a 1D lattice embedding a zero-dimensional (0D) chaotic 
scattering region instead of a long 1D localized region. 

Let us consider a single mode infinite lead embedding a scattering region invariant under time reversal and spin rotation symmetry. 
The scattering matrix $S$ is a $2\times2$ unitary symmetric matrix which we assume to be totally random at the Fermi energy 
$E_F$ (i. e. $S(E_F)$ is taken from the Circular Orthogonal Ensemble (COE)~\cite{mehta}). If the scattering region is made of a 
quantum dot forming a cavity, such an assumption can be justified if the corresponding classical trajectories are chaotic~\cite{doron,chaos}. 
If $S~\in~COE$, the distributions of its eigenvalues~\cite{mehta} $e^{i\theta_j}$ and of the transmission~\cite{jpb,jp,mb} read
\begin{equation} 
P({\theta_1},{\theta_2})=\frac{|e^{i\theta_1}-e^{i\theta_2}|}{16 \pi}; \  \  \ P(T)=\frac{1}{2\sqrt{T}}.
\label{COE-distribution}
\end{equation}
The eigenvalues of the matrix $Q$ -- the so-called delay-times $\tau_j$ -- also have a universal distribution for chaotic scattering: Assuming 
the WBL limit and a Gaussian scatterer of size $M \to \infty$, the inverse delay-times turn out to be given by the Laguerre 
ensemble from RMT~\cite{bfb1,bfb2} whereas, as usual in RMT ensembles, the eigenvectors of $Q$ are totally random and independent of its 
eigenvalues which are correlated. In particular, for a $2\times2$ matrix $Q$, the joint probability distribution $P({\tilde \tau}_1,{\tilde \tau}_2)$ 
of the two rescaled delay-times ${{\tilde \tau_j}}=\tau_j/\tau_H$ ($\tau_H=2\pi\hbar/\Delta_F$ being the Heisenberg time with $\Delta_F$ the mean level 
spacing of the scatterer at $E_F$) reads:
\begin{equation}
P({\tilde \tau}_1,{\tilde \tau}_2)=\frac{1}{48}|{\tilde \tau}_1-{\tilde \tau}_2| ({\tilde \tau}_1 {\tilde \tau}_2)^{-4} 
\exp [-\sum_{j=1}^2 \frac{1}{2{\tilde \tau}_j}].
\label{distribution-Time-delay-COE-Bulk}
\end{equation}
Integrating $P({\tilde \tau}_1,{\tilde \tau}_2)$ over one delay-time (see Appendix A), one gets for the average density of the 
rescaled delay-time: 
\begin{equation}
P_{B}({\tilde \tau})=\frac{(4{\tilde \tau}+1) \exp [-1/{\tilde \tau}] }{6{\tilde \tau}^4} - 
\frac{(4{\tilde \tau}-1)\exp [-1/(2{\tilde \tau})]}{12{\tilde \tau}^4}.    
\label{average-Time-delay-COE-Bulk}
\end{equation}
The distribution $P_{B}(\sigma_k)$ of the dimensionless thermopower 
\begin{equation}
\sigma_k\equiv \frac{\Delta_F}{2\pi}\frac{1}{T} \frac{dT}{dE}
\end{equation} 
is given in Ref.~\onlinecite{vlsb}: 
\begin{eqnarray}
P_{B}(\sigma_k)& = & \int\limits_{-1}^{+1} dc \int\limits_{0}^{\infty}d{\tilde \tau}_1\int\limits_{0}^{\infty} d{\tilde \tau}_2  
\int\limits_{0}^{+1} dT~f(c,{\tilde \tau}_1,{\tilde \tau}_2,T)\cr
&\times & \delta~\left(\sigma_k - c ({\tilde \tau}_1-{\tilde \tau}_2) \sqrt{\frac{1}{T}-1}\right)\,,
\label{Seebeck-COE-Bulk}
\end{eqnarray}
where $f(c,{\tilde \tau}_1,{\tilde \tau}_2,T)=\frac{1}{\pi \sqrt{1-c^2}}P(T)P({\tilde \tau}_1,{\tilde \tau}_2)F({\tilde \tau}_1,{\tilde \tau}_2)$. 
The charging effects are taken into account~\cite{vlsb,bvlfbb} by the function $F({\tilde \tau}_1,{\tilde \tau}_2)$: $C$ being the capacitance 
of the cavity, $F({\tilde \tau}_1,{\tilde \tau}_2)={\tilde \tau}_1+{\tilde \tau}_2$ if $e^2/C \gg \Delta_F$ while 
$F({\tilde \tau}_1,{\tilde \tau}_2)=1$ if $e^2/C \ll \Delta_F$. 
\begin{figure}
\includegraphics[keepaspectratio,width=\columnwidth]{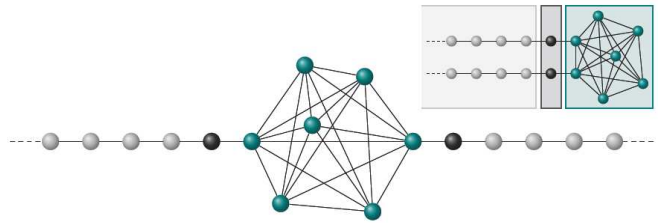}
\caption{\label{fig1} (Color online) Chaotic scatterer of $M=7$ sites (indicated in green) embedded in a 1D tight-binding lattice 
(gray sites with nearest-neighbor hopping $t=1$). The two end sites of the two leads are shown in a darker gray. Upper right: 
Partition of the same infinite system used for deriving Eqs.~(\ref{Scat1}\,-\,\ref{Scat3}).}
\end{figure}
The distributions (\ref{average-Time-delay-COE-Bulk}\,-\,\ref{Seebeck-COE-Bulk}) were obtained in the WBL limit, an approximation where the real 
part of the lead self energy $\Sigma(E)$ and the energy dependence of its imaginary part around $E_F$ are neglected. Moreover, $H_M$ was taken 
from the Gaussian orthogonal ensemble (GOE), a distribution giving rise to chaotic scattering only when $M\to \infty$ and $\Delta_F \to 0$. 
These assumptions, leading to distributions (\ref{average-Time-delay-COE-Bulk}\,-\,\ref{Seebeck-COE-Bulk}), cannot be made as $E_F$ approaches 
the edges of the conduction band and the spectrum edges of the chaotic scatterer. 

To study the delay-time and the thermopower distributions near the band edges, we consider a 1D lattice (hopping term $t$, lattice spacing $a=1$) embedding 
a scatterer of $M$ sites (see Fig.~\ref{fig1}).  $S(E_F)$ is calculated from the scatterer Hamiltonian $H_M$ and from the {\it exact} 
expressions of the lead self-energies $\Sigma(E)$.  The COE distribution for $S$ at $E_F$ is obtained~\cite{brouwerthesis} by taking for 
$H_M$ a Cauchy (Lorentzian) distribution of center $E_F/2$ and of width $\Gamma_F=t\sqrt{1-(E_F/2t)^2}$. This model gives rise to edge 
distributions which differ from the bulk distributions (\ref{average-Time-delay-COE-Bulk}\,-\,\ref{Seebeck-COE-Bulk}) and which turn out to 
be universal after an energy rescaling similar to the one used by Tracy and Widom for the energy levels~\cite{tracy-widom1,tracy-widom2,forrester}.

\section{Scattering outside the WideBand limit} 
Usually, the infinite system is divided into a scatterer and 
two attached leads, and the scattering matrix is given~\cite{datta} in terms of the scatterer Hamiltonian $H_M$ and of the lead self-energies $\Sigma(E)$. 
For the model sketched in Fig.~\ref{fig1}, it is more convenient when $M >2$ to divide the infinite system as indicated in 
the inset of Fig.~\ref{fig1}: a ``system'' made of the two sites (dark gray) located at the lead ends (Hamiltonian $H_0=V_0 {\bf 1}_2$), 
with the scatterer (green) at its right side and the two leads (gray) without their end sites at its left side. The scatterer 
and the leads are described by their self-energies~\cite{datta}. Using this partition, one obtains the scattering matrix $S$ at an energy $E$ in terms 
of an effective $2 \times 2$ energy-dependent ``Hamiltonian'' matrix ${\tilde H}_2(E)$:  
\begin{eqnarray}
S(E)&=&-{\bf 1}_2+ 2i \Gamma(E) A_2(E) \label{Scat1} \\ 
A_2(E)&=&\frac{1}{E{\bf 1}_2-H_0-{\tilde H}_2(E)-\Sigma(E)}\label{Scat2} \\
{\tilde H}_2(E)&=&W^{\dagger}\frac{1}{E{\bf 1}_M-H_M}W.\label{Scat3}
\end{eqnarray}
Here  ${\bf 1}_M$ is the $M \times M$ identity matrix and $W$ is an $M \times 2$ matrix with $W_{i\ne j}=0$, $W_{11}=W_{22}=t$.  
$E=-2t \cos k$, $\Sigma(E)=-t e^{ik}{\bf 1}_2$ is the lead self-energy and $\Gamma(E)=t \sin k = t \sqrt{1-(E/2t)^2}$.  
Since $V_0=0$, $H_0$ disappears from Eq.~(\ref{Scat2}).\\
\indent The scattering matrix $S$ obtained with this peculiar partition of the system is identical to the scattering matrix $\tilde{S}$ obtained 
with the standard partition, up to a phase factor $e^{2ik}$ since the ends of the two leads (shown in dark gray in the main panel of Fig.~\ref{fig1}) 
are included in the scattering region for the calculation of $S$ whereas they are excluded for $\tilde{S}$. Of course, this phase factor 
introduces a shift $\hbar/\Gamma$ of the delay-times, while it does not affect the transmission $T$ of the scatterer and its thermopower $S_k$ and 
\begin{equation}
S(E_F)~\in~COE \,\Leftrightarrow\, \tilde{S}(E_F)~\in~COE.
\end{equation}
In the following, we choose to define the delay-times for the scattering region containing the chaotic scatterer alone (\textit{i.e.} with respect 
to $\tilde{S}$) but in practice, for numerical reasons, we compute first the delay-times for $S$ and then shift them to deduce those for $\tilde{S}$.\\ 
\indent Hereafter, we determine how  ${\tilde H}_2(E_F)$ must be distributed to have $S(E_F)~\in~COE$. This will allow us to calculate the distributions of 
the (rescaled) delay-times $\tilde{\tau}_j$ and of $S_k$ at the edges of the conduction band.

\section{Chaotic Scattering and Cauchy Ensembles}
\label{section-2}
An $M \times M$ Hamiltonian $H_M$ has a Cauchy distribution 
$\mathcal{C}(M,{\epsilon},\Gamma)$ of center ${\epsilon}$ and width $\Gamma$ if the probability $P(dH_M)$ of finding $H_M$ 
inside the infinitesimal volume of measure $\mu(dH_M)=\prod_{i\leq j}^M dH_{M,ij}$ around $H_M$ is given~\cite{brouwerthesis} by 
\begin{equation}
P(dH_M)\propto \operatorname{det} \left((H_M-{\epsilon}{\bf 1}_M)^2+\Gamma^2 {\bf 1}_M \right)^{-\frac{M+1}{2}} \mu(dH_M).
\label{Cauchy}
\end{equation}

For $M=2$, it is simpler to use the usual expression~\cite{datta} giving the scattering matrix $\tilde{S}$ in terms of $H_2$
\begin{equation}
\tilde{S}(E)=-{\bf 1}_2+ \frac{2i \Gamma(E)}{E{\bf 1}_2-H_2-\Sigma(E)},
\label{Scat-direct}
\end{equation}
instead of using the previous partition leading to the matrix $S$ given by Eq.~(\ref{Scat1}). From Eq.~(\ref{Scat-direct}), 
one can see that $\tilde{S}$ can be diagonalized by the same orthogonal transformation (rotation of angle $\varphi$) which diagonalizes 
$H_2$. Therefore, the eigenvectors of $\tilde{S}$ and $H_2$ are identical and independent of $E$, while their eigenvalues $e^{i\theta_j}$ 
and $E_j$ are simply related. For an energy $E=E_F$, one gets
\begin{equation}
e^{i\theta_j}=-1+2i \frac{\Gamma_F}{E_F/2 - E_j+i\Gamma_F}.
\label{relation-S-H2}
\end{equation}
Chaotic scattering ($\tilde{S}(E_F)~\in~COE$) corresponds to a probability $P(d\tilde{S})=\frac{1}{V} \mu(d\tilde{S})$ of having $\tilde{S}$ at 
$E=E_F$ inside an infinitesimal volume of measure $\mu(d\tilde{S})$ ($V=\int \mu(d\tilde{S})$ being the total volume of the space where $\tilde{S}$ 
is defined). Therefore, the probability $P(dH_2)$ of the scatterer Hamiltonian $H_2$ necessary for making the scattering chaotic at $E_F$ is 
given by the condition $P(dH_2)= p(H_2) \mu(dH_2) =\frac{1}{V} \mu(d\tilde{S})$. Using the expressions giving the measures $\mu(d\tilde{S})$ and 
$\mu(dH_2)$ in terms of the eigenvalue-eigenvector coordinates~\cite{mehta} 
\begin{eqnarray}
&\mu(dH_2)=|E_1-E_2|dE_1 dE_2 d\varphi \\
&\mu(d\tilde{S})=|e^{i\theta_1}-e^{i\theta_2}|d\theta_1 d\theta_2 d\varphi,
\end{eqnarray}
and Eq.~(\ref{relation-S-H2}), one obtains the necessary and sufficient condition for having chaotic scattering at $E_F$:
\begin{equation}
S(E_F)~\in~COE \Leftrightarrow H_2~\in~{\mathcal C}(2,E_F/2,\Gamma_F).
\label{distribution-M=2}
\end{equation}

For $M>2$, the condition~(\ref{distribution-M=2}) for the effective Hamiltonian ${\tilde H}_2(E_F)$ instead of $H_2$ 
is sufficient and necessary for having $S(E_F)~\in~COE$. Moreover, one can use two properties of the Cauchy 
distributions~\cite{brouwerthesis},
\begin{eqnarray}
&H_M~\in~{\mathcal C}(M,\epsilon,\Gamma) \Rightarrow H_M^{-1}~\in~{\mathcal C}(M,\frac{\epsilon}{D},\frac{\Gamma}{D}) \nonumber \\
&H_M~\in~{\mathcal C}(M,\epsilon/D,\Gamma/D) \Rightarrow W^{\dagger} H_M W \in \mathcal{C}(2,\frac{t^2 \epsilon}{D},\frac{t^2\Gamma}{D}) \nonumber
\label{theorems}
\end{eqnarray}
where $D={\epsilon}^2+\Gamma^2$, to obtain the following result:  
\begin{equation}
H_M~\in~{\mathcal C}(M,E_F/2,\Gamma_F) \Rightarrow S(E_F)~\in~COE 
\label{distribution-M}
\end{equation}
since $t^2/D=1$ when $D=E_F^2/4+\Gamma_F^2$. 
While Eq.~(\ref{distribution-M=2}) is a necessary and sufficient condition, Eq.~(\ref{distribution-M}) is a sufficient condition 
which is \textit{a priori} not necessary.\\
\indent We point out that, contrary to the Gaussian ensemble, the Cauchy ensemble for $H_M$ allows for chaotic scattering (in the sense $S(E_F)~\in~COE$)
for any number $M \geq 2$. It opens the way to a \textit{scattering approach} even if the scattering region has a large but finite number $M$ of sites.
In this approach, the distribution of $S$ is independent of $E_F$ ($S(E_F)~\in~COE$ for chaotic scattering), while 
the distribution of the scatterer Hamiltonian becomes a function of $E_F$. This differs from an \textit{Hamiltonian approach}, where the probability 
of $H_M$ is independent of $E_F$, giving rise to a distribution of $S$ which depends on $E_F$ and may give rise to chaotic 
scattering only for a certain value of $E_F$. The scattering approaches are based on maximum entropy distributions for $S$, and not for 
$H_M$. The COE ensemble is a maximum entropy distribution for $S$. The Poisson kernel 
$(P(dS)\propto |\operatorname{det}(1-\left<S\right>^{\dagger}S)| \mu(dS)$ introduced in Ref.~\onlinecite{mpk} for statistical nuclear reactions 
is a maximum-entropy distribution of $S$ under the constraint that the ensemble average of $S$ is equal to a given matrix $\left<S\right>$. 
Similarly, the RMT description of disordered conductors introduced in Ref.~\onlinecite{mps} was based on a maximum entropy distribution for $S$ 
under the constraint of a given density of transmission eigenvalues. 

\section{ Scale Invariant Distributions at the Band Edges}
\label{edge distribution}
Hereafter, we study the distributions $P({\tilde \tau})$ and $P(\sigma_k)$ assuming that $H_M~\in~{\mathcal C}(M,E_F/2,\Gamma_F)$. 
In this section, we give the proof that these distributions become independent of the size $M \geq 2$ of the scattering region as 
$E_F$ approaches the band edges $\pm 2t$ of the leads. Therefore, to study a scattering region with $M=2$ sites only allows us to 
obtain the analytical expressions of these edge distributions. The proof is straightforward if one uses the expressions~(\ref{Scat1}-\ref{Scat3}) 
for $S$ instead of the standard ones for $\tilde{S}$. Indeed, we get from Eqs.~(\ref{Scat1}-\ref{Scat2}) 
\begin{eqnarray}
&\frac{\partial S}{\partial E}=2i\Gamma(E)\frac{\partial A_2}{\partial E}+2i\frac{\partial\Gamma}{\partial E} A_2(E) \\
&2i\Gamma\frac{\partial A_2}{\partial E}=-\frac{E}{2}A_2^2-2i\Gamma\left(\frac{A_2^2}{2}-A_2\frac{\partial {\tilde H}_2}{\partial E}A_2\right).
\label{2x2-edges}
\end{eqnarray}
As $E_F$ approaches the band edges $\pm 2t$ of the leads, $\Gamma_F \to 0$, the second term in the last equation can be neglected and so the 
distribution of $\partial S/\partial E$ at $E_F$ turns out to depend only on the distribution of the small $2 \times 2$  energy-dependent 
``Hamiltonian'' matrix ${\tilde H}_2(E_F)$, and not on the whole distribution of the large $M \times M$ Hamiltonian $H_M$. 
This implies that the distributions of $Q$ and of $S_k$  depend only on the distribution of the $2 \times 2$ matrix ${\tilde H}_2(E_F)$ at the 
band edges. Since we have shown that ${\tilde H}_2(E_F)~\in~{\mathcal C}(2,E_F/2,\Gamma_F)$ is a necessary and sufficient 
condition for chaotic scattering, independently of the size $M \geq 2$ of the scattering region, the distributions of the delay-time 
and of the thermopower (more generally of any function of $S$ and $\partial S/\partial E$) become independent of $M$ at the band 
edges if the condition $S(E_F \to \pm 2t)~\in~COE$ is imposed. Therefore, it is sufficient to study the case $M=2$ to obtain the edge 
distributions associated with chaotic scattering. The corresponding analytical expressions are given later in 
Eqs.~(\ref{Distributions-M=2-1},\ref{Distributions-M=2-2}) 
and derived in Appendix B. Let us proceed by showing now that the edges of the conduction band of the leads and the spectrum edges 
of a unique asymptotic Gaussian scatterer coincide if the scattering is chaotic. This result will give us the idea of introducing the relevant 
energy scale to use for measuring the distance between $E_F$ and the band edges $\pm 2t$.   

\begin{figure}
\includegraphics[keepaspectratio,width=6cm]{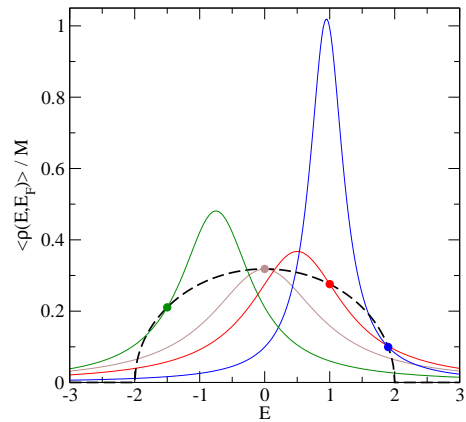}
\caption{\label{fig2}
(Color online) Level density per site (in units of $t^{-1}$), $\left\langle\rho(E,E_F)\right\rangle/M$, as a function of 
the energy $E$ (in units of $t$), for four Cauchy ensembles ($H_M \in {\mathcal C}(M,E_F/2,\Gamma_F)$), each of them 
giving rise to chaotic scattering at $E_F=0$ (brown line), $1$ (red line), $-1.5$ (green line) and $1.9$ (blue line). 
The circles indicate the value of the level density at $E=E_F$. The black dashed line is the semi-circle law 
$(\Delta_F M)^{-1}$ (see Eq.~(\ref{Density-GOE})).} 
\end{figure}
 
\section{ Chaotic Scattering and the Semi-Circle Law}

When $H_M~\in~{\mathcal C}(M,E_F/2,\Gamma_F)$, $S(E_F)~\in~COE$ and the average energy level density of $H_M$ at an energy $E$ and for a 
Fermi energy $E_F$ reads~\cite{brouwerthesis} 
\begin{equation}
\left\langle \rho(E,E_F) \right\rangle = \frac{M}{\pi t} \frac{\Gamma_F}{(E-E_F/2)^2+\Gamma_F^2} \\
\label{Density-GOE-E}
\end{equation}
This gives for $E=E_F$ 
\begin{equation}
\Delta_F^{-1}=\left\langle \rho(E=E_F,E_F) \right\rangle = \frac{1}{\delta}\sqrt{1-(E_F/2t)^2}, 
\label{Density-GOE}
\end{equation}
where $\delta=\frac{\pi t}{M}$ is the average level spacing at $E_F=0$. As shown in Fig.~\ref{fig2}, the average level density per site 
at $E_F$ of the scattering region must vary on a semi-circle as $E_F$ varies inside the conduction band for having $S(E_F)~\in~COE$. 
This semi-circle is also the limit when $M\to\infty$ of the level density of a {\it unique} asymptotic Gaussian ensemble where 
$\operatorname{tr} H_M^2$ has a zero average and a variance $V^2=2 \delta^2 M/\pi^2$. This establishes a very intriguing relation 
between a {\it single} Gaussian ensemble describing an infinite scattering region and a {\it family} of Cauchy ensembles 
$H_M~\in~{\mathcal C}(M,E_F/2,\Gamma_F)$ describing scattering regions of finite size. This family depends on a continuum parameter $E_F$ 
free to vary inside the conduction band. Another important conclusion can be drawn from Eq.~\eqref{Density-GOE}: Chaotic scattering with 
$\Delta_F^{-1}\to 0$ becomes possible if one of the energy distances $\epsilon_F^{\mp} \equiv E_F \pm 2t \to 0$. These limits correspond to 
the edges of the conduction band and to the spectral edges of the single Gaussian ensemble. This is particularly interesting, since the 
universality of the spectral fluctuation near the spectral edges is known for being different from that in the bulk (Tracy-Widom vs Wigner 
level distributions~\cite{tracy-widom1,tracy-widom2,forrester}). This makes likely that the distributions of $Q$ and $S_k$ should be different 
from the bulk distributions near the edges, and could give rise to a new asymptotic universality for the time-delay matrix $Q$ and for the 
thermopower $S_k$ when the Tracy-Widom scaling is adopted. 

\section{Energy rescaling near the band edges}
In our quest for universal edge distributions, let us introduce the dimensionless parameter 
\begin{equation}
\alpha \equiv   \frac{\Gamma_F^2}{\Delta_F t} = \frac{1}{8\pi} \left|\frac{\epsilon_F^{-}}{tM^{-1/3}}\right|^{3/2}\left|\frac{\epsilon_F^{+}}{tM^{-1/3}}\right|^{3/2}.
\label{alpha-definition}
\end{equation}
The reasons for introducing $\alpha$ (i. e. for measuring the energy distance from the edges in units of $t M^{-1/3}$) are twofold. 

First, $\alpha$ appears in the study of the case $M=2$. The calculations given in Appendix B yield the following expressions for 
the delay-time and thermopower distributions: 
\begin{eqnarray}
P_{M=2}({\bar \tau};\alpha)&=&\frac{4 \alpha}{\sqrt{1-(4\pi \alpha {\bar \tau})^2}} \label{Distributions-M=2-1}\\
P_{M=2}(\sigma_k;\alpha)&=&2 \alpha \ln \frac{1+\sqrt{1-(2 \pi \alpha\sigma_k)^2}}{2\pi \alpha |\sigma_k|},
\label{Distributions-M=2-2}
\end{eqnarray}
${\bar \tau} = {\tilde \tau} - \hbar/(2 \Gamma_F \tau_H)$ being a shifted delay-time. 
The distributions~(\ref{Distributions-M=2-1}) and~(\ref{Distributions-M=2-2}) are valid for $-2t \leq E_F \leq 2t$ if $M=2$. 
Moreover (see Sec.~\ref{edge distribution}), they describe also any Cauchy scatterer $H_M~\in~{\mathcal C}(M,E_F/2,\Gamma_F)$ 
of arbitrary size $M > 2$ in the limit $\alpha \to 0$ ($E_F \to \pm 2t$). 
 
Second, $\alpha$ is related near the band edges to the scale $x$ used at the spectrum edges of Gaussian matrices. In particular, for an 
Hamiltonian $H_M~\in~GOE$ having the semi-circle density shown in Fig.~\ref{fig2} when $M \to \infty$, the tail of the density 
$\left\langle\rho^{GOE}(E)\right\rangle/M$ takes the Tracy-Widom form outside the conduction band~\cite{vavilov}
\begin{equation}
\frac{\left\langle \rho^{GOE}(x)\right\rangle}{M}= \frac{1}{4 \sqrt{\pi} x^{1/4}} \exp [-\frac{2}{3} x^{3/2}],   
\label{Density-Tracy-Widom}
\end{equation}  
in terms of the scale $x=\epsilon/(tM^{-2/3})$. Since $\epsilon\equiv |E+2t|$  ($\epsilon \equiv |E-2t|$) if 
$E \approx -2t$ ($E \approx 2t$) respectively, $x \to (\pi \alpha)^{2/3}$ near the edges. Therefore, the 
distributions~(\ref{average-Time-delay-COE-Bulk}\,-\,\ref{Seebeck-COE-Bulk}) correspond to the limit where $x$ and hence 
$\alpha \to \infty$, since they were obtained in the limit where $\left\langle\rho^{GOE}(x)\right\rangle=0$ outside the band. This limit 
is out of reach if $M=2$ ($\alpha < 2/\pi$) and requires to consider $ M \gg 2$. 
\begin{figure}
\includegraphics[keepaspectratio,width=0.8\columnwidth]{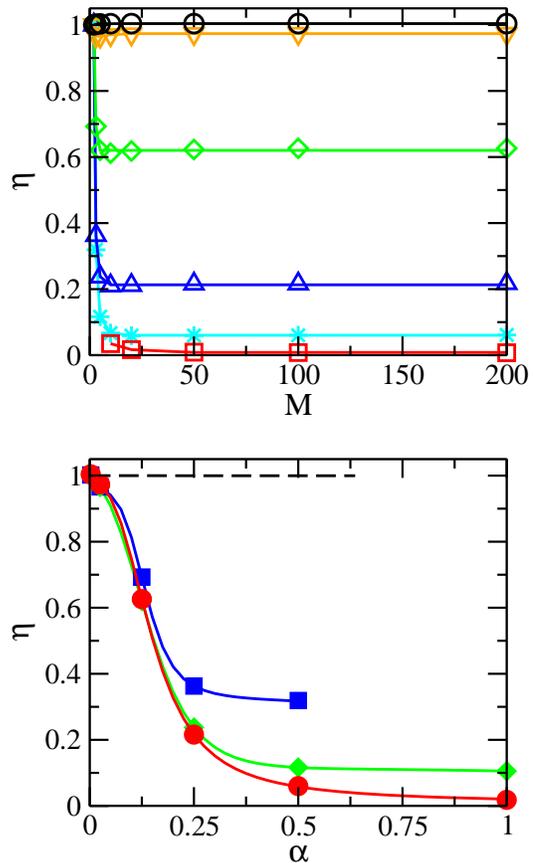}
\caption{\label{fig3} (Color online) Top: $\eta(\alpha,M)$ parameter (Eq.~\eqref{eta}) as a function of $M$ for $\alpha=0.0025$ 
({\Large$\circ$}), $0.025$ ({\color{orange}$\triangledown$}), $0.125$ ({\Large{\color{green}$\diamond$}}), $0.25$ ({\color{blue}$\vartriangle$}), 
$0.5$ ({\Large{\color{cyan}$\ast$}}) and $2.5$ ({\color{red}$\square$}). This shows that the 
thermopower distribution converges towards a $\alpha$-dependent asymptotic limit when $M\to\infty$. Bottom: $\eta(\alpha,M)$ parameter 
as a function of $\alpha$ for sizes $M=2$ (dashed line), $3$ ({\color{blue}$\blacksquare$}), $5$ ({\color{green}$\blacklozenge$}) and $100$ 
({\Large{\color{red}$\bullet$}}). The curve with $M=100$ gives a good approximation of the asymptotic crossover between the edge and the 
bulk behaviors of the thermopower distribution. In both panels, full lines are guides to the eye.}
\end{figure}

\begin{figure*}
\begin{center}
\includegraphics[keepaspectratio,width=0.85\linewidth]{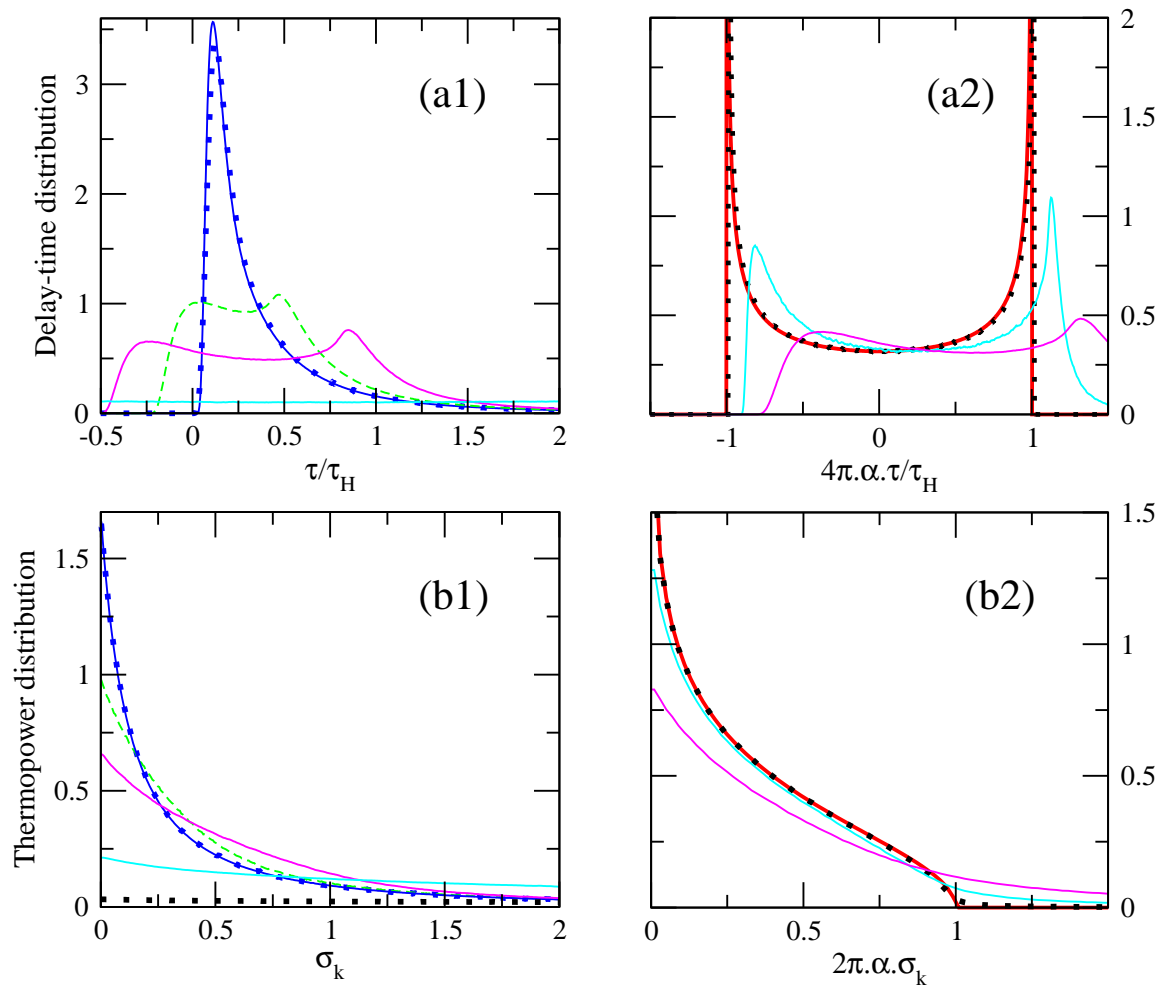}
\caption{
(Color online) Left panels: Asymptotic distributions of the delay-time $\tilde{\tau}=\tau/\tau_H$~(a1) and of the thermopower $\sigma_k$~(b1) for different values of $\alpha$. 
Data are shown for $\alpha=2.5\times 10^{-3}$ (black dotted line in~(b1)), 
$2.5\times 10^{-2}$ (cyan line), $0.125$ (magenta line), $0.25$ (green dashed line) and $2.5$ (blue dotted line). The data for $\alpha=2.5$ 
agree with the distributions (continuous blue line) expected in the limit $\alpha \to \infty$  
[Eqs.~(\ref{average-Time-delay-COE-Bulk}) and~(\ref{Seebeck-COE-Bulk}) for the delay-time and for the thermopower respectively]. 
Right panels: Same asymptotic distributions as a function of rescaled variables. 
The data for $\alpha=2.5\times 10^{-5}$ (black dotted line in~(a2)) and $\alpha=2.5\times 10^{-3}$ (black dotted line in~(b2)) 
agree with the distributions (continuous red line) expected in the limit $\alpha \to 0$ 
[Eqs.~(\ref{Distributions-M=2-1}\,-\,\ref{Distributions-M=2-2})].
}
\label{fig4}
\end{center}
\end{figure*}

\section{Universal $\alpha$-dependent asymptotic distributions} 

We study the distribution $P(\sigma_k;M,\alpha)$  as a function of the size $M$ of an Hamiltonian $H_M~\in~{\mathcal C}(M,E_F/2,\Gamma_F)$, 
taking $E_F$ closer to the nearest edge for keeping $\alpha$ constant. To measure this size dependence, we introduce the parameter
\begin{equation}
\eta(\alpha,M)=\frac{\int d\sigma_k |P(\sigma_k;M,\alpha)-P_{B}(\sigma_k)|}{\int d\sigma_k |P_{M=2}(\sigma_k;\alpha)-
P_{B}(\sigma_k)|}, 
\label{eta}
\end{equation} 
where $P_{B}(\sigma_k)$ [$P_{M=2}(\sigma_k;\alpha)$] is given by Eq.~(\ref{Seebeck-COE-Bulk}) [Eq.~(\ref{Distributions-M=2-2})] respectively. 
As shown in the upper panel of Fig.~\ref{fig3}, $P(\sigma_k;M,\alpha)$ does reach an asymptotic limit if $\alpha$ keeps a constant value. If 
$\alpha \approx 0$, $\eta \approx 1$ independently of $M$, confirming that $P_{M=2}(\sigma_k;\alpha)$ gives the asymptotic edge distribution. 
For intermediate values of $\alpha$, the size effects on $P(\sigma_k;M,\alpha)$ become quickly negligible as $M$ 
increases. In the lower panel of Fig~\ref{fig3}, $\eta(M)$ is given as a function of $\alpha$. In the asymptotic limit ($M \approx 100$), 
one can see a crossover around $\alpha \approx 0.25$ between the edge limit $\alpha \to 0$ and the bulk limit $\alpha \to \infty$. The same 
conclusion can be drawn from the study of $P({\tilde\tau};M,\alpha)$.\\
\indent In Fig.~\ref{fig4}, the asymptotic distributions (obtained with $M=200$) are given for different values of $\alpha$. The delay-time distributions are
shown in panels~(a1) and~(a2) and the thermopower distributions in panels~(b1) and~(b2), only for $\sigma_k >0$ since they are symmetrical around the origin. 
In the left panels~(a1) and~(b1) are also shown with blue lines the bulk behaviors, given by Eq.~(\ref{average-Time-delay-COE-Bulk}) for the delay-time and 
by the numerical integration of Eq.~(\ref{Seebeck-COE-Bulk}) for the thermopower, taking $F(\tilde{\tau_1},\tilde{\tau}_2)=1$. One can see that
the delay-time and thermopower distributions converge towards those of the bulk when $\alpha$ increases. This confirms that the 
distributions (\ref{average-Time-delay-COE-Bulk},\ref{Seebeck-COE-Bulk}) given in Refs.~\onlinecite{bfb1,vlsb} for chaotic cavities opened to leads via 
single mode point contacts characterize also a 1D lattice embedding a Cauchy scatterer as $\alpha\to\infty$. The convergence of the asymptotic distributions 
towards the edge limit $\alpha \to 0$ is shown in the right panels~(a2) and~(b2). We have multiplied the abscissas by $\alpha$ to get rid of the 
$\alpha$-dependency of the edge distributions~(\ref{Distributions-M=2-1}) and~(\ref{Distributions-M=2-2}) which are indicated with red lines in 
panels~(a2) and~(b2) respectively. Strictly speaking, we have plotted in~(a2) the distribution $P_{M=2}(\alpha\bar{\tau})$ in the limit $\alpha\to 0$, 
when its center goes to zero, $\alpha/\Gamma_F \to 0$. For very small values of $\alpha$, we find a perfect agreement between the distributions 
of large scatterers and the formula (\ref{Distributions-M=2-1}-\ref{Distributions-M=2-2}), evaluated in the limit $\alpha\to 0$ using a scatterer with 
$M=2$ sites only.\\ 
\indent The change of the delay-time distribution as a function of $\alpha$ is striking. When $\alpha \to \infty$, $\left\langle \tau 
\right\rangle \to \pi \hbar /\Delta_F$ and the fluctuations do not give rise to negative times. The scattering region is always attractive. 
When $\alpha \to 0$, $\left\langle \tau \right\rangle \to \hbar/(2 \Gamma_F)$ (see Eq.~(\ref{Distributions-M=2-1})). This change of average value 
induced by a decrease of $\alpha$ is similar to the one induced in compound nuclei by a decrease of the density of resonance levels~\cite{lyuboshitz}. 
Around this average, there are huge fluctuations which can yield positive or negative values for $\tau$, corresponding to an attractive or repulsive 
character of the scattering region. In the 1D localized limit studied in Ref.~\onlinecite{kumar}, a probability to observe negative values of $\tau$ occurs 
also at the band edges. Those negative delay-times occur in the limit where the Fermi velocity vanishes in the leads as the Fermi energy approaches the 
band edges, either in the 0D chaotic case studied here or in the 1D localized case studied in Ref.~\onlinecite{kumar}. 

\section{Summary and conclusions}

 We have studied the delay-time and the low ${\mathcal T}$ limit of the thermopower using large Cauchy scatterers and 1D leads. 
This corresponds to a scattering approach where an invariant distribution for $S$ is obtained from a continuum family of distributions for 
the scatterer Hamiltonian $H_M$, of arbitrary size $M$. We have shown that this family of Hamiltonian ensembles gives rise to a semi-circle law 
(with center and width given by the conduction band of the leads) for the average level density per site at $E_F$ of the scattering region. 
This intriguing coincidence between the average level density of a unique {\it asymptotic} Gaussian ensemble and the average level density 
at $E_F$ of a continuum family of Cauchy scatterers (describing scattering regions of finite size) has led us to use the same energy rescaling 
as the one leading to the Tracy-Widom expression at the GOE spectrum edges. Doing this, we have shown not only that the universal bulk distributions 
break down as $E_F$ approaches the edges of the conduction band (as previously noticed in the 1D localized limit), but also that a new asymptotic 
edge universality takes place when the Tracy-Widom energy rescaling is adopted. Signatures of this edge universality are given by the existence of 
asymptotic edge distributions for the delay-time and the thermopower which depend only on a single parameter $\alpha$. When $\alpha \to 0$, the 
delay-time and the thermopower have the same edge distributions for arbitrary $M$ as those for a $M=2$ scatterer, which we have obtained analytically. 
When $\alpha \to \infty$, the model used is described by the same universal bulk distributions which have been previously obtained assuming the wide band 
limit. 

This study raises an interesting question for the theory of random matrices: As a function of their location in the spectrum, the energy levels exhibit 
either Wigner-Dyson (bulk) or Tracy-Widom (edges) statistics. The fluctuating dependence of the energy levels on an external parameter in quantum systems 
with a chaotic classical dynamics has received considerable attention and is of fundamental importance for the theory of mesoscopic systems. These 
parametric correlations are understood in the bulk of the GOE spectra~\cite{as}, but to our knowledge not at 
their edges. We have shown that the time-delay matrix of our model (and hence the fluctuating dependence of the spectrum of $S$ on the energy $E$) 
exhibits a new edge universality. What are the parametric correlations at the spectrum edge? Our study suggests that they might also be universal after 
a suitable energy rescaling.

\acknowledgments{This research has been supported by CEA through the DSM-Energy Program (project NAT) and by the RTRA Triangle de la Physique 
(project Meso-Therm).} 

\appendix
\section{Average density of delay-times in the bulk of the conduction band} 
\label{app:dsttaubulk}
In this appendix, we explain how to get Eq.~(\ref{average-Time-delay-COE-Bulk}). We introduce the (rescaled) inverse delay-times 
$\tilde{\gamma}_j=1/\tilde{\tau}_j$ which are distributed according to the Laguerre ensemble~\onlinecite{bfb1,bfb2},
\begin{equation}
P(\tilde{\gamma}_1,\tilde{\gamma}_2)=\frac{1}{48}|\tilde{\gamma}_1-\tilde{\gamma}_2|\,\tilde{\gamma}_1\tilde{\gamma}_2\, 
e^{-(\frac{\tilde{\gamma}_1}{2}+\frac{\tilde{\gamma}_2}{2})}
\end{equation}
when the WBL limit is assumed. Their average density
\begin{equation}
P(\tilde{\gamma})=\frac{1}{2}\int_0^\infty \int_0^\infty d\tilde{\gamma}_1 d\tilde{\gamma}_2 
P(\tilde{\gamma}_1,\tilde{\gamma}_2)\sum_{j=1}^2\delta(\tilde{\gamma}-\tilde{\gamma}_j)
\end{equation}
is given in that case by Eqs.~(3.5)\,-\,(3.8) of Ref.~\cite{baker-forrester-pearce},
\begin{equation}
P_B(\tilde{\gamma})=\frac{1}{Z}\,\tilde{\gamma}\,e^{-\tilde{\gamma}/2}\,I(\tilde{\gamma})
\end{equation}
with $Z=48$ and $I(\tilde{\gamma})=4(4-\tilde{\gamma})-8(4+\tilde{\gamma})e^{-\tilde{\gamma}/2}$. Here we have kept the same notations as the ones 
used in Ref.~\onlinecite{baker-forrester-pearce} for the sake of clarity and we have added the subscript $B$ to $P_B$ in order to remind the reader that this result 
obtained within the WBL limit is valid only in the bulk of the conduction band. Eq.~(\ref{average-Time-delay-COE-Bulk}) for 
$P_B(\tilde{\tau})=P_B(\tilde{\gamma})|\frac{d\tilde{\gamma}}{d\tilde{\tau}}|$  (with $\tilde{\gamma}=1/\tilde{\tau}$) follows immediately.

\section{Study of the chaotic scatterer with $M=2$ sites}
\label{app:2sites}
In this appendix, we calculate analytically the average density of delay-times (Eq.~(\ref{Distributions-M=2-1})) and the thermopower 
distribution (Eq.~(\ref{Distributions-M=2-2})) of the two-sites cavity, assuming that the scattering matrix $\tilde{S}$ given by Eq.~(\ref{Scat-direct}) 
is always chaotic at the Fermi energy (\textit{i.e.} $\tilde{S}(E_F)~\in~\mathrm{COE}$ for any $E_F$). For both calculations, we exploit the fact that the 
(energy independent) rotation matrix $R_\varphi$ diagonalizing the Hamiltonian $H_2$ of the cavity,
\begin{equation}
R_\varphi = 
\begin{pmatrix}
\cos\varphi & -\sin\varphi \\
\sin\varphi & \cos\varphi
\end{pmatrix}          
\end{equation} 
also diagonalizes the scattering matrix, 
\begin{equation}
\label{eq_app:diagS}
\tilde{S}(E) = R_\varphi
\begin{pmatrix}
e^{i\theta_1(E)} & 0 \\
0 & e^{i\theta_2(E)}
\end{pmatrix}   
R_{-\varphi}  \,.     
\end{equation} 
Here, since $\tilde{S}(E_F)~\in~\mathrm{COE}$ by hypothesis, the real parameter $\varphi$ is uncorrelated with $\theta_1$ and $\theta_2$ and its 
distribution reads
\begin{equation}
P(\varphi)=\frac{1}{2\pi}\,,
\end{equation}
while the distribution of $\theta_1$ and $\theta_2$ at the Fermi energy is 
\begin{equation}
\label{eq_app:Pthetai}
P(\theta_1(E_F),\theta_2(E_F))=\frac{1}{16\pi}|e^{i\theta_1(E_F)}-e^{i\theta_2(E_F)}|\,.
\end{equation}

\subsection{Average density of delay-times}
\label{app:2sitestau}
By using Eq.~(\ref{Scat-direct}), it is straightforward to show that the time-delay matrix $\tilde{Q}=-i\hbar \tilde{S}^{-1}\partial \tilde{S}/ \partial E$ 
can be written as
\begin{equation}
\tilde{Q} = \frac{\hbar}{2\Gamma}\left[{\bf 1}_2+\left(\frac{1}{2}-i\frac{E}{4\Gamma}\right)\tilde{S}+\left(\frac{1}{2}+i\frac{E}{4\Gamma}\right)\tilde{S}^\dagger\right]\,.
\end{equation}
Thus we find, by inserting Eq.~(\ref{eq_app:diagS}), that $\tilde{Q}$ is also diagonalized by $R_\varphi$ and that its two eigenvalues -- the delay-times 
$\tau_i$ -- are
\begin{equation}
\label{eq_app:taui1}
\tau_i = \frac{\hbar}{2\Gamma}\left(1+\cos\theta_i+\frac{E}{2\Gamma}\sin\theta_i\right)\,.
\end{equation}
In the following, we evaluate them at the Fermi energy $E_F$ (even if not explicitly stated in the equations). For convenience, we re-write Eq.~(\ref{eq_app:taui1}) 
in the compact form
\begin{equation}
\bar{\tau}_i = \frac{1}{4\pi\alpha}\cos\bar{\theta}_i
\end{equation}
where $\bar{\tau}_i=(\tau_i-\frac{\hbar}{2\Gamma_F})/\tau_H$ is a rescaled and shifted delay-time, $\alpha$ is the parameter introduced in 
Eq.~(\ref{alpha-definition}) and $\bar{\theta}_i=\phi + \mathrm{sgn}(E_F)(\theta_i+\frac{\pi}{2})$ with $\phi=\arcsin\frac{\Gamma_F}{t}$. 
Thus, the average density of delay-times $P_{M=2}({\bar \tau})$ reads
\begin{equation}
\label{eq_app:intPtau}
P_{M=2}({\bar \tau})=\int_{-\pi}^{\pi}d\bar{\theta}\, P(\bar{\theta})\,\delta({\bar \tau}-\frac{1}{4\pi\alpha}\cos\bar{\theta})
\end{equation}
where $P(\bar{\theta})=\frac{1}{2\pi}$ is the average density of phases $\bar{\theta}_i$, obtained by integrating Eq.~(\ref{eq_app:Pthetai}) 
over one variable. We calculate the integral~(\ref{eq_app:intPtau}) by noticing first the even parity of the integrand and by making then the change 
of variable $u={\bar \tau}-\frac{1}{4\pi\alpha}\cos\bar{\theta}$. This leads to Eq.~(\ref{Distributions-M=2-1}).

\subsection{Thermopower distribution}
\label{app:2sitesSeebeck}
To calculate the thermopower distribution, the idea is to express the thermopower as a function of $\theta_1$ and $\theta_2$ only, since we know how 
those two variables are distributed at the Fermi energy. We start from Eq.~(\ref{eq_app:diagS}) which yields the following expression for the 
transmission $T$ of the scatterer,
\begin{equation}
T(E) = 4\cos^2(\varphi)\sin^2(\varphi)\sin^2\left(\frac{\theta_1-\theta_2}{2}\right)\,.
\end{equation}
To deduce the thermopower $S_k=\frac{1}{T}\frac{dT}{dE}$, we need to know in addition the energy derivatives $\frac{d\theta_i}{dE}$. By differentiating 
Eq.~(\ref{eq_app:taui1}), we find that they are given in our case by the delay-times, $\tau_i=\hbar\frac{d\theta_i}{dE}$, the parameter $\varphi$ being 
here energy-independent (in general, this is not the case). By using Eq.~(\ref{eq_app:taui1}), we finally get 
\begin{equation}
S_k(E) = -\frac{1}{\Gamma}\cos u\left(\sin v-\frac{E}{2\Gamma}\cos v\right)
\end{equation}
where $u=(\theta_1-\theta_2)/2$ and $v=(\theta_1+\theta_2)/2$. At the Fermi energy $E_F$, the joint distribution of the new variables $u$ and $v$ are given by 
Eq.~(\ref{eq_app:Pthetai}) up to a Jacobian factor, $P(u,v)=|\sin u|/(4\pi)$. Moreover, the rescaled thermopower $\sigma_k=\frac{\Delta_F}{2\pi}S_k$ 
can be written as
\begin{equation}
\sigma_k(E_F) = -\frac{1}{2\pi\alpha}\cos u \cos(\phi\pm v)
\end{equation}
with $\phi=\arcsin\frac{\Gamma_F}{t}$ and a $+$ [$-$] sign in the parentheses when $E_F>0$ [$E_F<0$]. Thus the thermopower distribution at the Fermi 
energy reads
\begin{equation}
P(\sigma_k)=\frac{1}{2}\int\limits_{-\pi}^{\pi}\int\limits_{-\pi}^{\pi}\! du\, dv P(u,v) \delta(\sigma_k+\frac{1}{2\pi\alpha}\cos u \cos(\phi\pm v))
\end{equation}
after extending the integration domain (hence the factor $1/2$ in front of the integral) by using the invariance of the integrand under the translations 
$u\to u \pm \pi$, $v\to v \pm \pi$. To calculate this integral, we make use first of the even parity in $u$ of the integrand. Second, we make the 
change of variable $u'=u-\frac{\pi}{2}$ and third the change of variable $p=\sin u'$. We get
\begin{equation}
\label{eq_app:PSk_2}
P(\sigma_k)=\frac{1}{4\pi}\int_{-1}^{1}\!\!dp\int_{-\pi}^{\pi}\!\!dv\, \delta(\sigma_k+\frac{1}{2\pi\alpha}p \cos(\phi\pm v))\,.
\end{equation}
The argument $F(v)$ of the delta function in Eq.~(\ref{eq_app:PSk_2}) vanishes at $v=v_0^\pm$ where $\cos(v_0^\pm \pm\phi)=-\frac{2\pi\alpha\sigma_k}{p}$. 
So by using the identity $\delta(F(v))=\sum_{\epsilon=\pm}\delta(v-v_0^\epsilon)|\frac{dF}{dv}(v_0^\epsilon)|^{-1}$ with 
$|\frac{dF}{dv}(v_0^\pm)|=\frac{1}{2\pi\alpha}\sqrt{p^2-(2\pi\alpha\sigma_k)^2}$, we get
\begin{equation}
\label{eq_app:Psk_3}
P(\sigma_k)=\frac{\alpha}{2}\int_{-1}^{1}\!\frac{dp}{\sqrt{p^2-(2\pi\alpha\sigma_k)^2}}\int_{-\pi}^{\pi}\!\!\!dv\,\sum_{\epsilon=\pm} \delta(v-v_0^\pm(p))\,.
\end{equation} 
When $2\pi\alpha|\sigma_k|>1$, the second integral in Eq.~(\ref{eq_app:Psk_3}) is zero and so $P(\sigma_k)=0$. When $2\pi\alpha|\sigma_k|<1$, we obtain
\begin{equation}
\label{eq_app:Psk_4}
P(\sigma_k)=2\alpha\int_{0}^{1}\!\frac{dp}{\sqrt{p^2-(2\pi\alpha\sigma_k)^2}}\Theta(p-2\pi\alpha|\sigma_k|)
\end{equation} 
where $\Theta$ is the Heaviside step function. Then, by making the change of variable $\cosh x = \frac{p}{2\pi\alpha|\sigma_k|}$, we get
\begin{equation}
P(\sigma_k)=2\alpha\cosh^{-1}\left(\frac{1}{2\pi\alpha|\sigma_k|}\right)\,,
\end{equation}
from which we deduce Eq.~(\ref{Distributions-M=2-2}) by using the identity $\cosh^{-1}(z) = \mathrm{ln}(z+\sqrt{z^2-1})$.

\bibliography{Paper}

\begin{thebibliography}{35}
\expandafter\ifx\csname natexlab\endcsname\relax\def\natexlab#1{#1}\fi
\expandafter\ifx\csname bibnamefont\endcsname\relax
  \def\bibnamefont#1{#1}\fi
\expandafter\ifx\csname bibfnamefont\endcsname\relax
  \def\bibfnamefont#1{#1}\fi
\expandafter\ifx\csname citenamefont\endcsname\relax
  \def\citenamefont#1{#1}\fi
\expandafter\ifx\csname url\endcsname\relax
  \def\url#1{\texttt{#1}}\fi
\expandafter\ifx\csname urlprefix\endcsname\relax\def\urlprefix{URL }\fi
\providecommand{\bibinfo}[2]{#2}
\providecommand{\eprint}[2][]{\url{#2}}

\bibitem[{\citenamefont{Godijn et~al.}(1999)\citenamefont{Godijn, Moller,
  Buhmann, Molenkamp, and van Langen}}]{molenkamp}
\bibinfo{author}{\bibfnamefont{S.~F.} \bibnamefont{Godijn}},
  \bibinfo{author}{\bibfnamefont{S.}~\bibnamefont{Moller}},
  \bibinfo{author}{\bibfnamefont{H.}~\bibnamefont{Buhmann}},
  \bibinfo{author}{\bibfnamefont{L.~W.} \bibnamefont{Molenkamp}},
  \bibnamefont{and} \bibinfo{author}{\bibfnamefont{S.~A.} \bibnamefont{van
  Langen}}, \bibinfo{journal}{Phys. Rev. Lett.} \textbf{\bibinfo{volume}{82}},
  \bibinfo{pages}{2927} (\bibinfo{year}{1999}).

\bibitem[{\citenamefont{van Langen et~al.}(1998)\citenamefont{van Langen,
  Silvestrov, and Beenakker}}]{vlsb}
\bibinfo{author}{\bibfnamefont{S.~A.} \bibnamefont{van Langen}},
  \bibinfo{author}{\bibfnamefont{P.~G.} \bibnamefont{Silvestrov}},
  \bibnamefont{and} \bibinfo{author}{\bibfnamefont{C.~W.~J.}
  \bibnamefont{Beenakker}}, \bibinfo{journal}{Supperlattices Microstruct.}
  \textbf{\bibinfo{volume}{23}}, \bibinfo{pages}{691} (\bibinfo{year}{1998}).

\bibitem[{\citenamefont{Sivan and Imry}(1986)}]{sivan-imry}
\bibinfo{author}{\bibfnamefont{U.}~\bibnamefont{Sivan}} \bibnamefont{and}
  \bibinfo{author}{\bibfnamefont{Y.}~\bibnamefont{Imry}},
  \bibinfo{journal}{Phys. Rev. B} \textbf{\bibinfo{volume}{33}},
  \bibinfo{pages}{551} (\bibinfo{year}{1986}).

\bibitem[{\citenamefont{van Houten et~al.}(1992)\citenamefont{van Houten,
  Molenkamp, Beenakker, and Foxon}}]{van-houten}
\bibinfo{author}{\bibfnamefont{H.}~\bibnamefont{van Houten}},
  \bibinfo{author}{\bibfnamefont{L.~W.} \bibnamefont{Molenkamp}},
  \bibinfo{author}{\bibfnamefont{C.~W.~J.} \bibnamefont{Beenakker}},
  \bibnamefont{and} \bibinfo{author}{\bibfnamefont{C.~T.} \bibnamefont{Foxon}},
  \bibinfo{journal}{Semicond. Sci. Technol.} \textbf{\bibinfo{volume}{7}},
  \bibinfo{pages}{215} (\bibinfo{year}{1992}).

\bibitem[{\citenamefont{Lunde and Flensberg}(2005)}]{lunde}
\bibinfo{author}{\bibfnamefont{A.~M.} \bibnamefont{Lunde}} \bibnamefont{and}
  \bibinfo{author}{\bibfnamefont{K.}~\bibnamefont{Flensberg}},
  \bibinfo{journal}{J. Phys. :Condens. Matter} \textbf{\bibinfo{volume}{17}},
  \bibinfo{pages}{3879} (\bibinfo{year}{2005}).

\bibitem[{\citenamefont{Vavilov and Stone}(2005)}]{stone-vavilov}
\bibinfo{author}{\bibfnamefont{M.~G.} \bibnamefont{Vavilov}} \bibnamefont{and}
  \bibinfo{author}{\bibfnamefont{A.~D.} \bibnamefont{Stone}},
  \bibinfo{journal}{Phys. Rev. B} \textbf{\bibinfo{volume}{72}},
  \bibinfo{pages}{205107} (\bibinfo{year}{2005}).

\bibitem[{\citenamefont{Balachandran et~al.}(2012)\citenamefont{Balachandran,
  Bosisio, and Benenti}}]{bosisio}
\bibinfo{author}{\bibfnamefont{V.}~\bibnamefont{Balachandran}},
  \bibinfo{author}{\bibfnamefont{R.}~\bibnamefont{Bosisio}}, \bibnamefont{and}
  \bibinfo{author}{\bibfnamefont{G.}~\bibnamefont{Benenti}},
  \bibinfo{journal}{Phys. Rev. B} \textbf{\bibinfo{volume}{86}},
  \bibinfo{pages}{035433} (\bibinfo{year}{2012}).

\bibitem[{\citenamefont{Smith}(1960)}]{smith}
\bibinfo{author}{\bibfnamefont{F.~T.} \bibnamefont{Smith}},
  \bibinfo{journal}{Phys. Rev.} \textbf{\bibinfo{volume}{118}},
  \bibinfo{pages}{349} (\bibinfo{year}{1960}).

\bibitem[{\citenamefont{Brouwer
  et~al.}(1997{\natexlab{a}})\citenamefont{Brouwer, Frahm, and
  Beenakker}}]{bfb1}
\bibinfo{author}{\bibfnamefont{P.~W.} \bibnamefont{Brouwer}},
  \bibinfo{author}{\bibfnamefont{K.~M.} \bibnamefont{Frahm}}, \bibnamefont{and}
  \bibinfo{author}{\bibfnamefont{C.~W.~J.} \bibnamefont{Beenakker}},
  \bibinfo{journal}{Phys. Rev. Lett.} \textbf{\bibinfo{volume}{78}},
  \bibinfo{pages}{4737} (\bibinfo{year}{1997}{\natexlab{a}}).

\bibitem[{\citenamefont{Brouwer et~al.}(1999)\citenamefont{Brouwer, Frahm, and
  Beenakker}}]{bfb2}
\bibinfo{author}{\bibfnamefont{P.~W.} \bibnamefont{Brouwer}},
  \bibinfo{author}{\bibfnamefont{K.~M.} \bibnamefont{Frahm}}, \bibnamefont{and}
  \bibinfo{author}{\bibfnamefont{C.~W.~J.} \bibnamefont{Beenakker}},
  \bibinfo{journal}{Waves in Random Media (special issue on disordered electron
  systems} \textbf{\bibinfo{volume}{9}}, \bibinfo{pages}{91}
  (\bibinfo{year}{1999}).

\bibitem[{\citenamefont{Tracy and Widom}(1994)}]{tracy-widom1}
\bibinfo{author}{\bibfnamefont{C.~A.} \bibnamefont{Tracy}} \bibnamefont{and}
  \bibinfo{author}{\bibfnamefont{H.}~\bibnamefont{Widom}},
  \bibinfo{journal}{Commun. Math. Phys.} \textbf{\bibinfo{volume}{159}},
  \bibinfo{pages}{151} (\bibinfo{year}{1994}).

\bibitem[{\citenamefont{Tracy and Widom}(1996)}]{tracy-widom2}
\bibinfo{author}{\bibfnamefont{C.~A.} \bibnamefont{Tracy}} \bibnamefont{and}
  \bibinfo{author}{\bibfnamefont{H.}~\bibnamefont{Widom}},
  \bibinfo{journal}{Commun. Math. Phys.} \textbf{\bibinfo{volume}{177}},
  \bibinfo{pages}{727} (\bibinfo{year}{1996}).

\bibitem[{\citenamefont{Forrester}(1993)}]{forrester}
\bibinfo{author}{\bibfnamefont{P.~J.} \bibnamefont{Forrester}},
  \bibinfo{journal}{Nucl. Phys. B} \textbf{\bibinfo{volume}{402}},
  \bibinfo{pages}{709} (\bibinfo{year}{1993}).

\bibitem[{\citenamefont{Doron et~al.}(1990)\citenamefont{Doron, Smilansky, and
  Frenkel}}]{doron}
\bibinfo{author}{\bibfnamefont{E.}~\bibnamefont{Doron}},
  \bibinfo{author}{\bibfnamefont{U.}~\bibnamefont{Smilansky}},
  \bibnamefont{and} \bibinfo{author}{\bibfnamefont{A.}~\bibnamefont{Frenkel}},
  \bibinfo{journal}{Phys. Rev. Lett.} \textbf{\bibinfo{volume}{65}},
  \bibinfo{pages}{3072} (\bibinfo{year}{1990}).

\bibitem[{\citenamefont{Genack et~al.}(1999)\citenamefont{Genack, Sebbah,
  Stoytchev, and van Tiggelen}}]{tiggelen}
\bibinfo{author}{\bibfnamefont{A.~Z.} \bibnamefont{Genack}},
  \bibinfo{author}{\bibfnamefont{P.}~\bibnamefont{Sebbah}},
  \bibinfo{author}{\bibfnamefont{M.}~\bibnamefont{Stoytchev}},
  \bibnamefont{and} \bibinfo{author}{\bibfnamefont{B.~A.} \bibnamefont{van
  Tiggelen}}, \bibinfo{journal}{Phys. Rev. Lett.}
  \textbf{\bibinfo{volume}{82}}, \bibinfo{pages}{715} (\bibinfo{year}{1999}).

\bibitem[{\citenamefont{Schanze et~al.}(2005)\citenamefont{Schanze, Stockmann,
  Martinez-Mares, and Lewenkopf}}]{schanze}
\bibinfo{author}{\bibfnamefont{H.}~\bibnamefont{Schanze}},
  \bibinfo{author}{\bibfnamefont{H.~J.} \bibnamefont{Stockmann}},
  \bibinfo{author}{\bibfnamefont{M.}~\bibnamefont{Martinez-Mares}},
  \bibnamefont{and} \bibinfo{author}{\bibfnamefont{C.~H.}
  \bibnamefont{Lewenkopf}}, \bibinfo{journal}{Phys. Rev. E}
  \textbf{\bibinfo{volume}{71}}, \bibinfo{pages}{016223}
  (\bibinfo{year}{2005}).

\bibitem[{\citenamefont{Gopar et~al.}(1996)\citenamefont{Gopar, Mello, and
  Buttiker}}]{gmb}
\bibinfo{author}{\bibfnamefont{V.~A.} \bibnamefont{Gopar}},
  \bibinfo{author}{\bibfnamefont{P.~A.} \bibnamefont{Mello}}, \bibnamefont{and}
  \bibinfo{author}{\bibfnamefont{M.}~\bibnamefont{Buttiker}},
  \bibinfo{journal}{Phys. Rev. Lett.} \textbf{\bibinfo{volume}{77}},
  \bibinfo{pages}{3005} (\bibinfo{year}{1996}).

\bibitem[{\citenamefont{Brouwer and Buttiker}(1997)}]{bb}
\bibinfo{author}{\bibfnamefont{P.~W.} \bibnamefont{Brouwer}} \bibnamefont{and}
  \bibinfo{author}{\bibfnamefont{M.}~\bibnamefont{Buttiker}},
  \bibinfo{journal}{Europhys. Lett.} \textbf{\bibinfo{volume}{37}},
  \bibinfo{pages}{441} (\bibinfo{year}{1997}).

\bibitem[{\citenamefont{Ringel et~al.}(2008)\citenamefont{Ringel, Imry, and
  Entin-Wohlman}}]{rie}
\bibinfo{author}{\bibfnamefont{Z.}~\bibnamefont{Ringel}},
  \bibinfo{author}{\bibfnamefont{Y.}~\bibnamefont{Imry}}, \bibnamefont{and}
  \bibinfo{author}{\bibfnamefont{O.}~\bibnamefont{Entin-Wohlman}},
  \bibinfo{journal}{Phys. Rev. B} \textbf{\bibinfo{volume}{78}},
  \bibinfo{pages}{165304} (\bibinfo{year}{2008}).

\bibitem[{\citenamefont{Texier and Comtet}(1999)}]{texier-comtet}
\bibinfo{author}{\bibfnamefont{C.}~\bibnamefont{Texier}} \bibnamefont{and}
  \bibinfo{author}{\bibfnamefont{A.}~\bibnamefont{Comtet}},
  \bibinfo{journal}{Phys. Rev. Lett.} \textbf{\bibinfo{volume}{82}},
  \bibinfo{pages}{4220} (\bibinfo{year}{1999}).

\bibitem[{\citenamefont{Anantha~Ramakrishna and Kumar}(2001)}]{kumar}
\bibinfo{author}{\bibfnamefont{S.}~\bibnamefont{Anantha~Ramakrishna}}
  \bibnamefont{and} \bibinfo{author}{\bibfnamefont{N.}~\bibnamefont{Kumar}},
  \bibinfo{journal}{Eur. Phys. J. B} \textbf{\bibinfo{volume}{23}},
  \bibinfo{pages}{515} (\bibinfo{year}{2001}).

\bibitem[{\citenamefont{Mehta}(1991)}]{mehta}
\bibinfo{author}{\bibfnamefont{M.~L.} \bibnamefont{Mehta}},
  \emph{\bibinfo{title}{{Random Matrices 2nd edn}}} (\bibinfo{publisher}{New
  York: Academic}, \bibinfo{year}{1991}).

\bibitem[{\citenamefont{Giannoni et~al.}(1991)\citenamefont{Giannoni, Voros,
  and Zinn-Justin}}]{chaos}
\bibinfo{author}{\bibfnamefont{M.-J.} \bibnamefont{Giannoni}},
  \bibinfo{author}{\bibfnamefont{A.}~\bibnamefont{Voros}}, \bibnamefont{and}
  \bibinfo{author}{\bibfnamefont{J.}~\bibnamefont{Zinn-Justin}},
  \emph{\bibinfo{title}{{Chaos and Quantum Physics}}}
  (\bibinfo{publisher}{North-Holland, Amsterdam}, \bibinfo{year}{1991}).

\bibitem[{\citenamefont{Jalabert et~al.}(1994)\citenamefont{Jalabert, Pichard,
  and Beenakker}}]{jpb}
\bibinfo{author}{\bibfnamefont{R.~A.} \bibnamefont{Jalabert}},
  \bibinfo{author}{\bibfnamefont{J.~L.} \bibnamefont{Pichard}},
  \bibnamefont{and} \bibinfo{author}{\bibfnamefont{C.~W.~J.}
  \bibnamefont{Beenakker}}, \bibinfo{journal}{Europhys. Lett.}
  \textbf{\bibinfo{volume}{27}}, \bibinfo{pages}{255} (\bibinfo{year}{1994}).

\bibitem[{\citenamefont{Jalabert and Pichard}(1995)}]{jp}
\bibinfo{author}{\bibfnamefont{R.~A.} \bibnamefont{Jalabert}} \bibnamefont{and}
  \bibinfo{author}{\bibfnamefont{J.~L.} \bibnamefont{Pichard}},
  \bibinfo{journal}{J. Phys. I France} \textbf{\bibinfo{volume}{5}},
  \bibinfo{pages}{287} (\bibinfo{year}{1995}).

\bibitem[{\citenamefont{Baranger and Mello}(1994)}]{mb}
\bibinfo{author}{\bibfnamefont{H.~U.} \bibnamefont{Baranger}} \bibnamefont{and}
  \bibinfo{author}{\bibfnamefont{P.~A.} \bibnamefont{Mello}},
  \bibinfo{journal}{Phys. Rev. Lett.} \textbf{\bibinfo{volume}{73}},
  \bibinfo{pages}{142} (\bibinfo{year}{1994}).

\bibitem[{\citenamefont{Brouwer
  et~al.}(1997{\natexlab{b}})\citenamefont{Brouwer, van Langen, Frahm,
  Buttiker, and Beenakker}}]{bvlfbb}
\bibinfo{author}{\bibfnamefont{P.~W.} \bibnamefont{Brouwer}},
  \bibinfo{author}{\bibfnamefont{S.~A.} \bibnamefont{van Langen}},
  \bibinfo{author}{\bibfnamefont{K.~M.} \bibnamefont{Frahm}},
  \bibinfo{author}{\bibfnamefont{M.}~\bibnamefont{Buttiker}}, \bibnamefont{and}
  \bibinfo{author}{\bibfnamefont{C.~W.~J.} \bibnamefont{Beenakker}},
  \bibinfo{journal}{Phys. Rev. Lett.} \textbf{\bibinfo{volume}{79}},
  \bibinfo{pages}{913} (\bibinfo{year}{1997}{\natexlab{b}}).

\bibitem[{\citenamefont{Brouwer}(1997)}]{brouwerthesis}
\bibinfo{author}{\bibfnamefont{P.~W.} \bibnamefont{Brouwer}},
  \bibinfo{journal}{Ph.D thesis, Leiden University}  (\bibinfo{year}{1997}).

\bibitem[{\citenamefont{Datta}(1995)}]{datta}
\bibinfo{author}{\bibfnamefont{S.}~\bibnamefont{Datta}},
  \emph{\bibinfo{title}{{Electronic Transport in Mesoscopic Systems}}}
  (\bibinfo{publisher}{Cambridge University Press}, \bibinfo{year}{1995}).

\bibitem[{\citenamefont{Mello et~al.}(1985)\citenamefont{Mello, Pereyra, and
  Seligman}}]{mpk}
\bibinfo{author}{\bibfnamefont{P.~A.} \bibnamefont{Mello}},
  \bibinfo{author}{\bibfnamefont{P.}~\bibnamefont{Pereyra}}, \bibnamefont{and}
  \bibinfo{author}{\bibfnamefont{T.~H.} \bibnamefont{Seligman}},
  \bibinfo{journal}{Ann. Phys. (N. Y.)} \textbf{\bibinfo{volume}{161}},
  \bibinfo{pages}{254} (\bibinfo{year}{1985}).

\bibitem[{\citenamefont{Muttalib et~al.}(1987)\citenamefont{Muttalib, Pichard,
  and Stone}}]{mps}
\bibinfo{author}{\bibfnamefont{K.~A.} \bibnamefont{Muttalib}},
  \bibinfo{author}{\bibfnamefont{J.-L.} \bibnamefont{Pichard}},
  \bibnamefont{and} \bibinfo{author}{\bibfnamefont{A.~D.} \bibnamefont{Stone}},
  \bibinfo{journal}{Phys. Rev. Lett.} \textbf{\bibinfo{volume}{59}},
  \bibinfo{pages}{2475} (\bibinfo{year}{1987}).

\bibitem[{\citenamefont{Vavilov et~al.}(2001)\citenamefont{Vavilov, Brouwer,
  Ambegaokar, and Beenakker}}]{vavilov}
\bibinfo{author}{\bibfnamefont{M.~G.} \bibnamefont{Vavilov}},
  \bibinfo{author}{\bibfnamefont{P.~W.} \bibnamefont{Brouwer}},
  \bibinfo{author}{\bibfnamefont{V.}~\bibnamefont{Ambegaokar}},
  \bibnamefont{and} \bibinfo{author}{\bibfnamefont{C.~W.~J.}
  \bibnamefont{Beenakker}}, \bibinfo{journal}{Phys. Rev. Lett.}
  \textbf{\bibinfo{volume}{86}}, \bibinfo{pages}{874} (\bibinfo{year}{2001}).

\bibitem[{\citenamefont{Lyuboshitz}(1978)}]{lyuboshitz}
\bibinfo{author}{\bibfnamefont{V.~L.} \bibnamefont{Lyuboshitz}},
  \bibinfo{journal}{JETP Lett.} \textbf{\bibinfo{volume}{28}},
  \bibinfo{pages}{30} (\bibinfo{year}{1978}).

\bibitem[{\citenamefont{Altshuler and Simons}(1995)}]{as}
\bibinfo{author}{\bibfnamefont{B.~L.} \bibnamefont{Altshuler}}
  \bibnamefont{and} \bibinfo{author}{\bibfnamefont{B.~D.}
  \bibnamefont{Simons}}, in \emph{\bibinfo{booktitle}{Mesoscopic Quantum
  Physics}}, edited by
  \bibinfo{editor}{\bibfnamefont{E.}~\bibnamefont{Akkermans}},
  \bibinfo{editor}{\bibfnamefont{G.}~\bibnamefont{Montambaux}},
  \bibinfo{editor}{\bibfnamefont{J.-L.} \bibnamefont{Pichard}},
  \bibnamefont{and}
  \bibinfo{editor}{\bibfnamefont{J.}~\bibnamefont{Zinn-Justin}}
  (\bibinfo{publisher}{North Holland, Amsterdam}, \bibinfo{year}{1995}).

\bibitem[{\citenamefont{Baker et~al.}(1998)\citenamefont{Baker, Forrester, and
  Pearce}}]{baker-forrester-pearce}
\bibinfo{author}{\bibfnamefont{T.}~\bibnamefont{Baker}},
  \bibinfo{author}{\bibfnamefont{P.}~\bibnamefont{Forrester}},
  \bibnamefont{and} \bibinfo{author}{\bibfnamefont{P.}~\bibnamefont{Pearce}},
  \bibinfo{journal}{J. Phys. A} \textbf{\bibinfo{volume}{31}},
  \bibinfo{pages}{6087} (\bibinfo{year}{1998}).

\end{thebibliography}

\end{document}